\documentclass[a4paper,aps,pre,twocolumn,superscriptaddress,groupedaddress]{revtex4}  
\usepackage[margin=1in]{geometry}
\usepackage{eufrak}
\usepackage{blkarray}
\def\doubleunderline#1{\underline{\underline{#1}}}
\usepackage{natbib}
\usepackage{graphicx}  
\usepackage{makecell}
\usepackage{dcolumn}   
\usepackage{bm}        
\usepackage{amssymb,amsmath}   
\usepackage{array,float}
\usepackage{color}
\newcommand{\nn}{\nonumber}
\newcommand{\be}{\begin{equation}} \newcommand{\ee}{\end{equation}}
\newcommand{\ba}{\begin{eqnarray}} \newcommand{\ea}{\end{eqnarray}}
\def\bg#1\eg{%
    \begin{gather}%
    #1%
    \end{gather}%
}
\newcommand{\vect}[1]{\boldsymbol{#1}}

\newcommand{\eq}[1]{\mbox{Eq. (\ref{eq:#1})}}
\newcommand{\fig}[1]{\mbox{Fig.~\ref{fig:#1}}}

\def\showjon{} 
\ifdefined\showjon
\newcommand{\jon}[1]{\color{blue}  #1  \color{black}}
\else
\newcommand{\jon}[1]{}
\fi

\newcommand{\qd}{\,.}
\newcommand{\qc}{\, ,}
\newcommand{\III}{\text{I/II}}
\newcommand{\I}{\text{I}}
\newcommand{\II}{\text{II}}

\makeatletter
\newcommand*{\balancecolsandclearpage}{%
  \close@column@grid
  \clearpage
  \twocolumngrid
}
\makeatother

\begin{document}
\title{Energy and Matter Supply for Active Droplets}
\author{Jonathan Bauermann}
\affiliation{Max Planck Institute for the Physics of Complex Systems, N\"othnitzer Stra\ss e 38, 01187 Dresden, Germany}

\author{Christoph A. Weber}
\email[]{christoph.weber@physik.uni-augsburg.de}
\affiliation{Faculty of Mathematics, Natural Sciences, and Materials Engineering: Institute of Physics, University of Augsburg, Universit\"atsstra{\ss}e\ 1, 86159 Augsburg, Germany}

\author{Frank J\"ulicher}
\email[]{julicher@pks.mpg.de}
\affiliation{Max Planck Institute for the Physics of Complex Systems, N\"othnitzer Stra\ss e 38, 01187 Dresden, Germany}
\affiliation{Center for Systems Biology Dresden, Pfotenhauerstrasse 108, 01307 Dresden, Germany}
\affiliation{Cluster of Excellence Physics of Life, TU Dresden, 01062 Dresden, Germany}
\date{\today}

\begin{abstract}
Chemically active droplets provide simple models for cell-like systems that can grow and divide.
Such active droplet systems are driven away from thermodynamic equilibrium and turn over chemically, which corresponds to a simple metabolism.
We consider two scenarios of non-equilibrium driving.
First, droplets are driven via the system boundaries by external reservoirs that supply nutrient and remove waste (boundary-driven).
Second, droplets are driven by a chemical energy provided by a fuel in the bulk (bulk-driven).
For both scenarios, we discuss the conservation of energy and matter as well as the balance of entropy.
We use conserved and non-conserved fields to analyze the   non-equilibrium steady states of active droplets.
Using an effective droplet model, we explore droplet stability and instabilities leading to droplet division.
Our work reveals that droplet division occurs quite generally in active droplet systems.
Our results suggest that life-like processes such as metabolism and division can emerge in simple non-equilibrium systems that combine the physics of phase separation and chemical reactions.
\end{abstract}
\maketitle


\section{Introduction}

Chemical processes in living cells are organised in distinct compartments. Some of these compartments use membranes, others use interfaces that separate coexisting phases~\cite{hyman2014liquid,shin2017liquid, banani:2017,boeynaems:2018}.
An important question is how compartmentalization provided by droplets affects chemical processes.
It has already been proposed hundred years ago that chemical compartmentalization by phase-separated coacervates could have played a key role at the origin of life~\cite{haldane:1929,oparin:1953,oparin:1959, hyman2012retrospect}.
In particular, phase-separated droplets could serve as simple models of protocells with life-like behaviours when maintained away from equilibrium:
(i) Droplets enrich and confine certain components providing a distinct chemical environment.
(ii) Such droplets can localise chemical reactions defining a proto-metabolism.
(iii) Chemical activity can drive droplet growth.
(iv) Chemically active droplets can undergo a shape instability, and divide into daughter droplets.
The life-life behaviors (i)-(iv) have been discussed theoretically using minimal models as a proof of principle~\cite{rashevsky:1934,zwicker:2017, weber2019physics}.
Such minimal models, however, consider simple binary mixtures that are composed of two components which undergo demixing and also can chemically convert into each other -- a simplification that can be hardly realized in experimental chemical system.

In this work, we discuss more realistic scenarios of droplet models of protocells.
These scenarios distinguish between nutrient, droplet material, waste and solvent.
Such systems are versatile and open new possibilities for experimental realisations since real chemical systems with active droplets~\cite{spath2021molecular,abbas2021peptide, schwarz2021parasitic, nakashima2021active} are not binary and the spatial distribution of nutrients and waste is in general relevant for the emergence of the aforementioned life-like behaviors.
Droplet systems also provide appealing models of simple but realistic protocells and permit to consider both energy and matter flows.

The paper is organised as follows. In Section~\ref{sect:models}, we first introduce  two complementary cases of active droplets which provide models of protocells which differ in the supply of matter and energy.
In Section~\ref{sect:theory_active_drops}, we present a general theory of droplet systems in the presence of chemical reactions.
We use this approach in Section~\ref{eq:protocell_dyanmics} to discuss the dynamics of protocells for the two cases.
In Section\ref{sect:energymassentropy_balance}, we discuss the energy, mass and entropy balance in these systems and we discuss our results in Section~\ref{sect:discussion}.

\section{Active droplet models of simple protocells}\label{sect:models}

We introduce active droplet models based on four components, $i=S,D,N,W$.
These components are solvent $S$, droplet material $D$, nutrient $N$ and waste $W$.
We first specify the components involved in phase separation and introduce the chemical reactions. We then discuss the conditions by which the system is driven out of thermodynamic equilibrium.

\subsection{Phase separation and chemical reactions}
\label{Section:II}

\begin{figure}[b]
  \centering
    \includegraphics[width=0.49\textwidth]{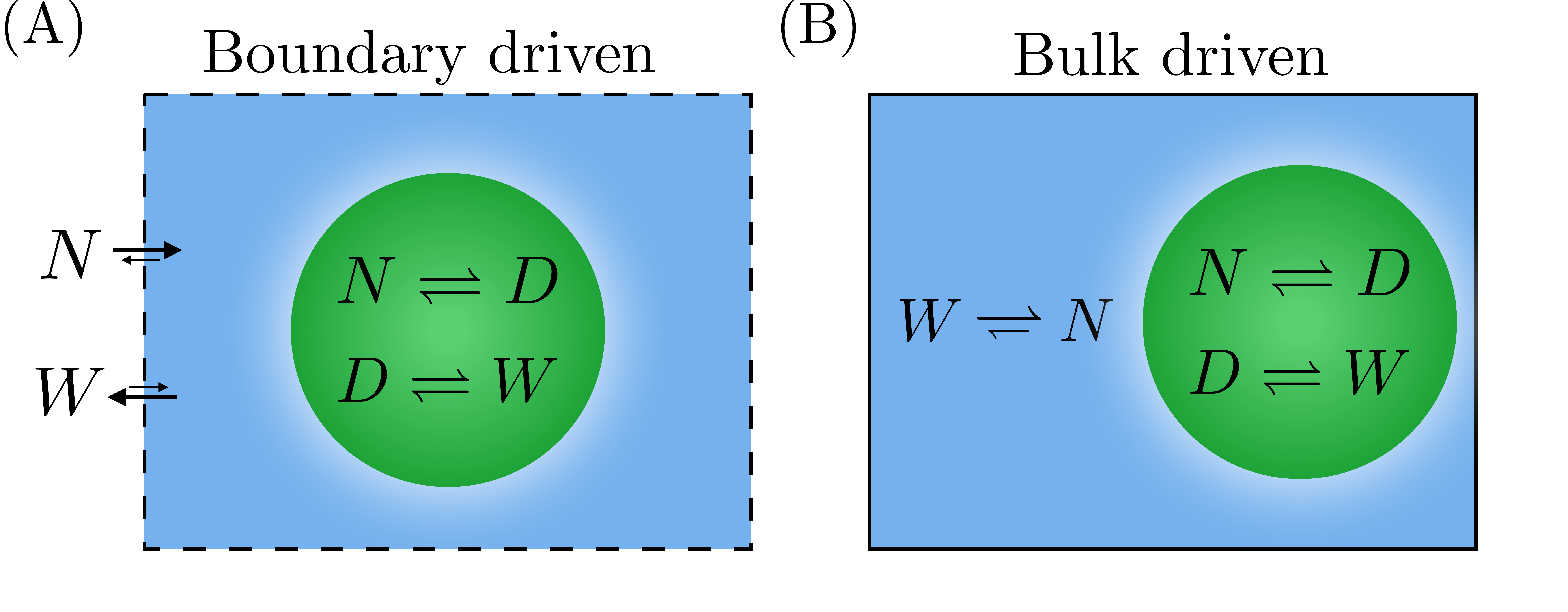}
    \label{fig:models_sketch}
    \caption{
    \textbf{Chemical reactions maintained away from equilibrium by reservoirs of nutrient and waste.}
    (A): Boundary-driven case with  nutrient and waste reservoirs at the system boundary.
    (B): Bulk-driven case with energy supply via chemical reactions with broken detailed balance in the outside phase.
    }
    \label{fig:schemes}
\end{figure}

We consider an incompressible, phase-separating system that undergoes chemical reactions, see Fig.~\ref{fig:schemes}.
Droplet material $D$ phase-separates from a solvent $S$ leading to
droplets.
A nutrient component $N$ can chemically convert to droplet material $D$, thereby feeding the droplet.  Droplet material can undergo a chemical change to become waste $W$. These reactions can be written as
\bg
N \rightleftharpoons D \; \;\; \;(\alpha=1) \qc \label{eq:reac_1}\\
D \rightleftharpoons W \; \;\; \;(\alpha=2) \qc
\label{eq:reac_2}
\eg
and are indexed with $\alpha=1,2$.
For simplicity, we consider that reactions involving the droplet material only occur inside the droplets.
This is motivated by the idea that these reactions are enabled by catalysts that are absent outside and thus solely located inside droplets.
We also consider the possibility that waste $W$ can be directly converted to nutrient $N$.
This reaction can be written as
\be
W \rightleftharpoons N \; \;\; \;(\alpha=3) \qd \label{eq:reac_3}
\ee
This reaction, for simplicity, only occurs outside of the droplet.

\subsection{Energy and Matter supply}

In order to discuss the energy balance, we need to introduce the exchange chemical potentials $\bar{\mu}_D$, $\bar{\mu}_N$ and $\bar{\mu}_W$ with respect to the solvent component.
We discuss two different cases of non-equilibrium driving, boundary-driven and bulk-driven, see \fig{schemes}(A,B).

In the boundary driven case (A), the concentrations of nutrient $N$, waste $W$ and droplet material $D$ at the system boundary are fixed via concentration boundary conditions, corresponding to a coupling to a reservoir.
This reservoir set values of the chemical potentials at the boundary $\bar{\mu}_i^\infty$. We choose conditions where $\bar{\mu}_N^\infty>\bar{\mu}_D^\infty>\bar{\mu}_W^\infty$ such that the direction of chemical reactions is from nutrient via droplet material to waste. Therefore, nutrient will be provided and waste will be absorbed by the reservoir.
In this case, we do not consider the direct reaction between waste and nutrient \eq{reac_3}.
In this case (A), the free energy provided by the reservoir per consumed nutrient molecule is $\bar{\mu}_N^\infty-\bar{\mu}_W^\infty>0$. This difference maintains the system out-of-equilibrium and drives a constant flux of molecules through the system.
The total volume fraction $\psi=\phi_D+\phi_N+\phi_W$, where $\phi_i$ denotes the volume fraction of component $i$, is a conserved quantity. It obeys the continuity equation, $\partial_t \psi + \nabla \cdot \vect{j}_\psi =0$, where $\vect{j}_\psi$ is the associated, conserved current.
The conserved quantity is provided by the reservoir and can feed droplet growth. Therefore, droplet size is not limited by the conservation law.

In the bulk driven case (B), we impose no-flux boundary conditions, considering that no exchange of molecules with reservoirs occurs at the boundaries.
The system is maintained out-of-equilibrium by introducing an external energy input $\Delta \mu^\text{act}$ to the reaction \eq{reac_3} ($\alpha=3$). This energy input could, for example, be provided by chemical fuel or by radiation.
The energy input $\Delta \mu^\text{act}$ drives the reaction which would spontaneously run from $N$ to $W$ in the opposite direction and allows nutrient to be recycled to waste.
This can be achieved when $\Delta \mu^\text{act} + \bar{\mu}_W>\bar{\mu}_N$.
The quantity $\psi$ is again a conserved quantity.
Since there is no exchange at the boundaries,
the amount of this conserved quantity $\psi$ is fixed inside the system. Therefore, the amount of droplet material $D$ is limited by the conserved quantity. This is implied that even though free energy $\Delta \mu^\text{act}$ is continuously supplied to the system, the droplet size is limited by the conservation law.

\subsection{Droplet dynamics and droplet division}

\begin{figure*}[tb]
    \includegraphics[width=0.8\textwidth]{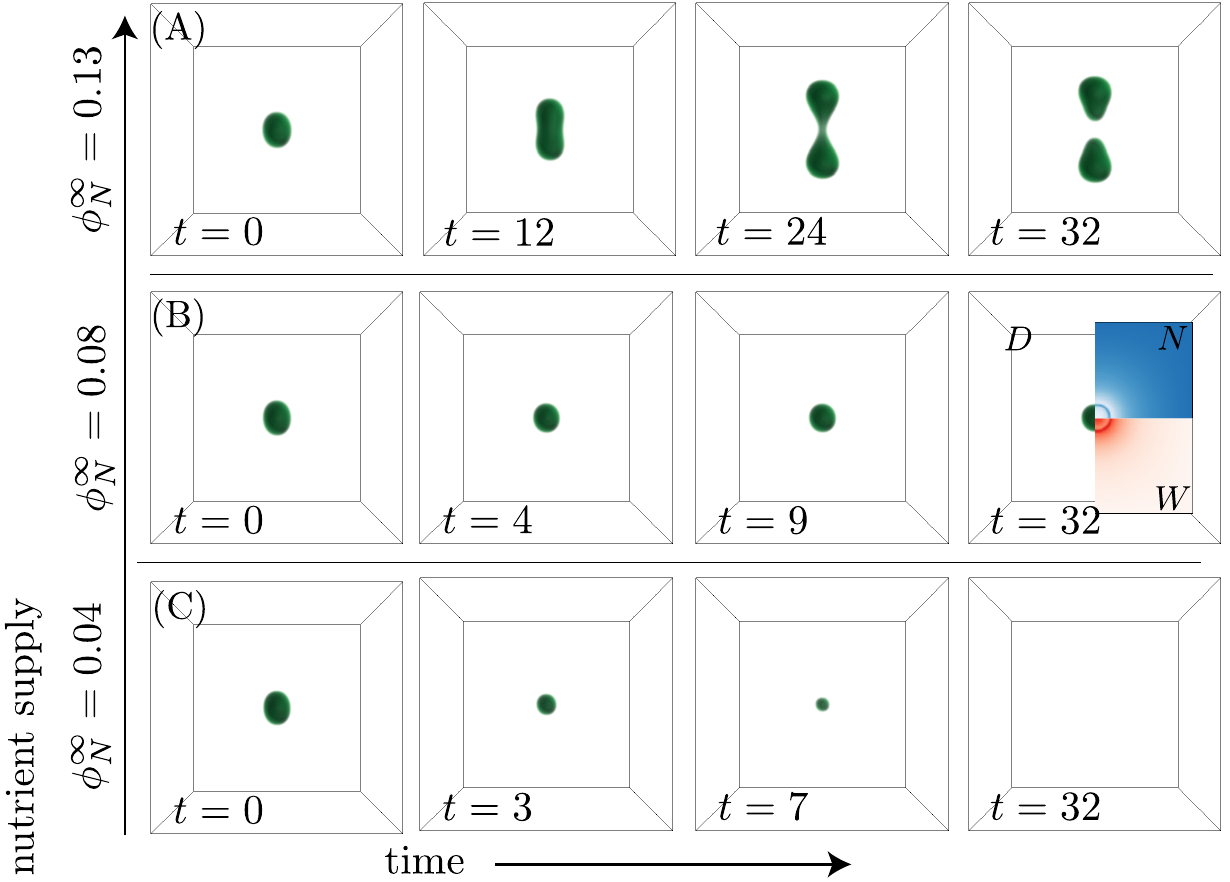} 
    \caption{
    \textbf{Shape dynamics of active droplets.}
    (A-C) Volume fraction fields of droplet material $\phi_D$ for different time points t in boundary-driven systems for three
    values of $\phi_N$ as indicated. Last panel in (B) shows density distributions of $\phi_N$ and $\phi_W$ on a planar cross section.
    Solutions were  obtained by numerical integration of Eq. (4), initializing with a slightly elongated droplet at t = 0.
    Parameter values are given in Table~1 and Appendix~\ref{app:parameters}.}
    \label{fig:num}
\end{figure*}

These cases can be compared to the minimal protocell model based on a binary system studied previously~\cite{zwicker:2017}.
In this simple systems, two components phase-separate from each other and can also be chemical converted into each other.
Active droplets in this binary system are driven in the bulk by the chemical energy input $\Delta \mu^\text{act}$. However, in this binary system, there is only one independent volume fraction and no conservation law. Therefore,
droplet size is not limited by a conservation law.
In previous work~\cite{zwicker:2017}, it was shown that droplets can either shrink and disappear, grow until droplets reach a stable size or undergo a shape instability and divide. In the latter case, cycles of growth and division can lead to many daughter droplets that together occupy an increasing volume.

Motivated by these findings, we study here the dynamics and properties of active droplets in the more realistic protocell models introduced here.
These models differ from the binary model by taking into account a solvent $S$ which phase-separates from the droplet material $D$. Most importantly, this solvent does not undergo chemical transitions with the droplet material $D$.
Therefore, the proposed models in our work are closer to existing experimental systems~\cite{spath2021molecular,abbas2021peptide, schwarz2021parasitic, nakashima2021active}.

As we will show in the following, our simple protocell models also exhibit regimes were droplets are stable with a finite size, and regimes where droplets divide via a shape instability.
Examples of droplet dynamics in the boundary driven case are shown in Fig.~\ref{fig:num} as snapshots of configurations of droplet material at different time points and for different nutrient reservoir volume fractions $\phi_N^\infty$.
For small nutrient supply ($\phi_N^\infty=0.04$), an initially prepared droplet tends to shrink as the waste release from the droplets exceeds the nutrient supply. For larger nutrient supply ($\phi_N^\infty=0.08$), an active droplet is stationary with a size that is determined by the nutrient supply. Increasing the nutrient supply further ($\phi_N^\infty=0.13$), the droplet shape becomes unstable and division occurs.

In the next chapter we present the general theory of multi-component active droplets and discuss in detail the energetics and the conservation laws.

\section{Theory of chemically active droplets}\label{sect:theory_active_drops}

The dynamics of active droplets is governed by chemical potentials~\cite{weber2019physics}. These chemical potentials determine both the thermodynamics of phase separation and the thermodynamics of chemical reactions.

\subsection{Irreversible thermodynamics of phase separation and reactions}

We consider an incompressible mixture of $M+1$ chemical components, denoted as $C_i$, with $i=0,...,M$. The local composition is given by the volume fractions $\phi_i$ with $i=1,...,M$ indicating the different components.
The volume fraction $\phi_0$, which we identify with the solvent $S$, is not an independent variable, because of the constraint $\sum_{i=0}^M \phi_i=1$.
The volume fractions are related to the local concentrations $n_i$ via $\phi_i = \nu_i n_i$, where $\nu_i$ is the molecular volume of components $i$. Incompressibility implies that the molecular volumes $\nu_i$ are constant parameters at constant temperature.
The dynamical equations for the volume fractions read
\be
\partial_t \phi_i = -  \vect{\nabla} \cdot \vect{j}_i  + r_i \qc \label{eq:dynamic}
\ee
where $\vect{j}_i$ is a diffusive current of volume fraction and $r_i$ denote source and sink terms resulting from chemical reactions.
Incompressibility requires that chemical reactions conserve volume, which implies $\sum_{i=0}^M r_i = 0$.

The diffusive flux arises due to spatial gradients of the exchange chemical potentials $\mu_i = \nu_i \delta F/ \delta \phi_i$, where $F[\phi_1, ...,\phi_M]$ is the free energy and $ \delta F/ \delta \phi_i$ denotes a functional derivative. In general, we can write
\be
\mu_i=k_B T \log \left( \gamma_i \phi_i \right) + \omega_i - \kappa_i \nabla^2 \phi_i \qc
\label{eq:chem_pot}
\ee
where $(\gamma_i \phi_i)$ is the exchange activity of components $i$ relative to the solvent, $\gamma_i$ is the corresponding activity coefficient and $\omega_i$ are the reference chemical potentials.
For mean-field models up to the second viral coefficients, the activity coefficient has the form
\be
 {\gamma}_i=\phi_0^{-\frac{\nu_i}{\nu_0}}\exp\left(\nu_i \frac{\sum_{j=1}^M  (\chi_{ij}-\chi_{i0}-\chi_{j0}) \phi_j }{k_B T} \right) \qc
 \label{eq:act_coef}
\ee
where the matrix $\chi_{ij}$ describes the molecular interactions, see Appendix~\ref{app:free_energy}.
The contribution $\omega_k$ is independent of composition and corresponds to a reference chemical potential.
Furthermore, $k_B$ is the Boltzmann constant and $T$ temperature. The coefficient $\kappa_i$ characterizes the free energy contributions due to gradients of composition and is related to interfacial tension.

The diffusive flux of volume fraction is driven by chemical potential gradients
\be
\vect{j}_i = -\nu_i \sum_{j=1}^M \Lambda_{ij} \vect{\nabla} \mu_j \qc
\ee
where $\Lambda_{ij}$ denotes the mobility matrix with $\Lambda_{ij}=\Lambda_{ji}$. For simplicity, we choose $\Lambda_{ij} =\lambda \phi_i (\delta_{ij}-\phi_j)$, where $\lambda$ is a molecular mobility. The diffusion matrix is then given by
\be
D_{ij}=\nu_i \sum_{k=1}^M  \Lambda_{ik} \frac{\partial \mu_k(\phi)}{ \partial\phi_j} \qc \label{eq:diff_const}
\ee
which becomes diagonal $D_{ij}=k_B T \lambda \, \delta_{ij}$ for vanishing molecular interactions and all molecular volumes being equal, $\nu_i=\nu$.
Thus, all molecular components diffusive independently of each other with the same diffusion coefficient.

The source rate $r_i$ of component $i$ in Eq.~\eqref{eq:dynamic} stems from chemical reactions.
Each chemical reaction $\alpha=1,...,L$ can be written as
\be
\sum_{i=0}^M \sigma_{i \alpha}^{+} C_i \underset{r_\alpha^{-}}{\stackrel{r_\alpha^{+}}{\rightleftharpoons}}  \sum_{i=0}^M \sigma_{i \alpha}^{-} C_i \qc
\ee
where $C_i$ is the chemical symbol of components $i$ and $\sigma_{i \alpha}^{\pm}$ are stoichiometric coefficients.
The net chemical rate per volume $r_\alpha = r_\alpha^{+}-r_\alpha^{-}$ can be written as a difference of the forward and backward rates.
We also define the stoichiometric matrix  $\sigma_{i \alpha} = \sigma_{i \alpha}^{-}-\sigma_{i \alpha}^{+}$ .

The relation between the source rates $r_i$ and the chemical rates $r_\alpha$ reads
\be
r_i = \sum_{\alpha=1}^S \nu_i \sigma_{i \alpha}  r_\alpha \qc
\label{eq:flux_gener}
\ee
and the stoichiometric coefficients $\sigma_{i \alpha}$ for volume conserving reactions must obey
\be \sum_{i=0}^M \nu_i \sigma_{i \alpha}  = 0 \qd \label{eq:volume_con}
\ee
The chemical reactions are driven by the reaction Gibbs free energy
\be
\Delta \mu_\alpha
= \sum_{i=1}^M \sigma_{i \alpha} \mu_i \qc
\ee
which can be expressed in terms of the exchange chemical potential $\mu_i$ given in Eq.~\eqref{eq:chem_pot}. The Gibbs free energy is the free energy change associated with a single event of reaction $\alpha$.
Microscopic reversibility imposes a detailed balance condition for forward and backward rates:
\be
\frac{r_\alpha^{+}}{r_\alpha^{-}} = \exp\left(- \frac{\Delta \mu_\alpha}{k_B T} \right) \qd \label{eq:detailed_balance}
\ee
This condition ensures that the system relaxes toward thermodynamic equilibrium in the absence of driving.
Chemical rates that satisfy this detailed balance condition can in general be written as
\be
r_\alpha^\pm = k_\alpha  \exp\left(\frac{\sum_{i=1}^M\sigma_{i \alpha}^\pm \mu_i}{k_B T}\right) \qc \label{eq:reactionrate_plusminus}
\ee
where $k_\alpha$ is a kinetic coefficient. The kinetics of the reactions is specified by the values and composition dependence of the kinetic coefficients $k_\alpha$.
We will use the freedom to choose this dependence of $k_\alpha$ on $\phi_i$ to localise reactions either inside or outside of droplets.

Substituting the exchange chemical potentials~\eqref{eq:chem_pot} in the relation of the source rate~\eqref{eq:flux_gener} gives
\begin{align}
\label{eq:reaction_rates}
&r_i=\sum_{\alpha=1}^S  k_\alpha \sigma_{i \alpha} \nu_i\\
& \times \left(\prod_{n=0}^M (\rho_n \phi_n)^{\sigma_{n\alpha}^+}
-
\prod_{n=0}^M (\rho_n\phi_n)^{\sigma_{n\alpha}^-}
\right),
\nonumber
\end{align}
where
\be \rho_n = \gamma_n \exp \left(\frac{\omega_n-\kappa_n \nabla^2 \phi_n
}{k_B T} \right) \qd
\ee
In dilute and homogeneous systems, $\rho_n$ are constant parameters, while in phase-separated systems, the coefficients $\rho_n$ depend on composition.



\subsection{Conserved and non-conserved densities}
The volume fractions $\phi_i$ are not conserved, which
follows from the presence of the source rates $r_i$ in the balance equation~\eqref{eq:dynamic}. In general, there are $L'\leq M$ linearly independent  chemical reactions. Therefore, there remain $C=M-L'$
conserved densities in addition to the conserved volume.
These conserved densities can be expressed as linear combinations of volume fractions.
We write the conserved densities as
\be \psi_j = \sum_{i=0}^M A_{ji} \phi_i \qc
\label{eq:conserved}\ee
where $j=0,1,...,C$. Here, the matrix $\doubleunderline{A}$ obeys the relation $\sum_{i=0}^M A_{ji} \nu_i \sigma_{i\alpha}=0$, i. e., the rows of the matrix $\doubleunderline{A}$ are linearly independent null-vectors of the matrix $\nu_i \sigma_{i \alpha}$. The conservation law \eq{volume_con} corresponds to $j=0$ with $A_{0i}=1/\nu_0$, which also clarifies that the units of $\psi_j$ are inverse volume.
We can also define the $L'=M-C$ non-conserved densities
\be \xi_\alpha = \sum_{0=1}^M B_{\alpha i} \phi_i \qc \label{eq:nonconserved} \ee
where $\alpha=1,...,L'$, which are the extents of reactions.
The reaction extents $\xi_\alpha$ measure the cumulative number of reaction events that have occurred per volume.
Here, the densities $\psi_j$ and $\xi_\alpha$ obey
\begin{subequations}
\bg
\partial_t \psi_j = -\nabla \cdot \vect{j}^\psi_j  \label{eq:dyn_con}\qc\\
\partial_t \xi_\alpha =-\nabla \cdot \vect{j}^\xi_\alpha+ r_\alpha \qc \label{eq:dyn_ex}
\eg
\end{subequations}
where we introduce the currents $\vect{j}^\psi_j = \sum_{i=0}^M A_{ji} \vect{j}_i$ of the conserved densities and the extent currents
and $\vect{j}^\xi_\alpha = \sum_{i=0}^M B_{\alpha i} \vect{j}_i$. Note that volume conservation corresponds to $\psi_0 = \sum_{i=0}^M \phi_i/\nu_0 = 1/\nu_0$ is not a dynamical variable and $j_0^\psi=0$.

The matrices $\doubleunderline{A}$ and $\doubleunderline{B}$ define the matrix elements of an invertible square matrix
\be
\doubleunderline{U}=
\left(
    \begin{array}{ccccc}
       A_{00 }&\cdots&A_{0i}&\cdots&A_{0M}\\
       &\cdots&A_{ji}&\cdots&\\
       A_{C0 }&\cdots&A_{Ci}&\cdots&A_{CM}\\

       B_{1 i}&...&B_{1i}&...&B_{1M}\\
       &...&B_{\alpha i}&...&\\
       B_{L' 0}&...&B_{L'i}&...&B_{L'M}
    \end{array}
\right)
\label{eq:Umatrix} \qd
\ee
A property of the inverse matrix $\doubleunderline{U}^{-1}$  is  $\doubleunderline{U}^{-1}_{i\alpha}= \nu_i \sigma_{i \alpha}$.
Note that the matrix $\doubleunderline{A}$ and the conserved densities $\psi_j$ are not unique. Any linear combination of conserved densities is also a conserved density. A choice of linearly independent conserved densities specifies the matrix $\doubleunderline{A}$ uniquely.
Similarly, the matrix $\doubleunderline{B}$ is also not unique, because each reaction extent is defined with respect to a reference value.

From the conserved densities $\psi_j$ and the reaction extents $\xi_\alpha$, we can obtain the volume fractions,
\be
\phi_i = \sum_{j=0}^{C}U^{-1}_{ij} \psi_j + \sum_{\alpha=1}^{L'}\nu_i \sigma_{i\alpha} \xi_\alpha \qd
\ee

\subsection{Energetics and entropy production}

We now discuss the energetics and thermodynamics of chemically active droplets.
The free energy of the system, $F=\int d^3x \, f$, from which we derive the chemical potential~\eqref{eq:chem_pot}, can be expressed in terms of the free energy density $f$.
The density of internal energy, $e=f+Ts$, can be decomposed
into the free energy density and an entropy density
$s=-\partial f/ \partial T$.
Energy conservation can be written as
\be
\partial_t e + \nabla \cdot \vect{j}_h = -\nabla \cdot \vect{j}_q \qc
\label{eq:firstlaw}
\ee
where we decompose the flux of energy into a heat flux $\vect{j}_q$ and an enthalpy flux $\vect{j}_h$. Note that \eq{firstlaw} defines the heat flux.
For simplicity, we focus on isothermal systems at constant temperature $T$.
The heat flux is determined by the constraint of fixed temperature, which corresponds to the limit of large heat conductance.

The flux of enthalpy is given as $\vect{j}_h = \sum_{i=1}^M  h_i  \vect{j}_i /\nu_i$, where we define the enthalpy $h_i = \mu_i + T s_i$
per molecule $C_i$ with $s_i = -\partial \mu_i /\partial T$ denoting the entropy per molecule.

The rate of change of internal energy density reads $\partial_t e = \sum_{i=1}^M h_i \partial_t \phi_i/\nu_i$.
Using \eq{firstlaw}, we can identify the heat production rate as
\be
\nabla \cdot \vect{j}_q = -\sum_{\alpha=1}^{L'}r_\alpha \Delta h_\alpha - \sum_{i=1}^M  \frac{\vect{j}_i}{\nu_i} \cdot \nabla h_i \qc \label{eq:heat}
\ee
where $\Delta h_\alpha =\sum_{i=1}^M \sigma_{i\alpha} h_i$ is the reaction enthalpy.
Using the definitions of the entropy density $s$ and energy density $e$, we obtain the entropy balance
\be
\partial_t s + \nabla \cdot \vect{j}_s =  \dot{\Theta}  \qc 
\ee
where the entropy flux is $\vect{j}_s= \sum_{i=1}^M s_i \vect{j}_i /\nu_i +  \vect{j}_q /T$.
The entropy production rate $\dot{\Theta}$ obeys
\bg
T \dot{\Theta} = -\sum_{\alpha=1}^{L'}r_\alpha \Delta \mu_\alpha -  \sum_{i=1}^M \frac{\vect{j}_i}{\nu_i} \cdot \nabla \mu_i  \qc  \label{eq:entropy}
\eg
which is zero or positive according  to the second law of thermodynamics.

In summary, we have derived a complete description for phase-separated systems in the presence of chemical reactions at constant  temperature. The system is captured by $C$ conserved quantities $\psi$, and $L'$ reaction extents $\xi$, described by \eq{dyn_con}  and \eq{dyn_ex} and obeys the energy conservation, as well as heat and entropy production are described by \eq{heat} and \eq{entropy}.
Note that volume changes of coexisting phases are associated with conserved fluxes $\vect{j}^\psi$.




\section{Dynamics of protocells}\label{eq:protocell_dyanmics}

In this section we specify the particular model introduced in Section~\ref{Section:II} and study the dynamics of active droplets in the boundary and bulk driven cases, respectively.

\subsection{Model Parameters}
Molecular interactions are described by the matrix with elements $\chi_{ij}$.
We consider $M=3$ non-solvent components $D$, $N$ and $W$ as well a solvent $S$.
For simplicity, all molecular components have the same molecular volume $\nu_i=\nu$.
The choice of interactions is based on the idea that droplet material phase-separates from the solvent.
The matrix is symmetric and the diagonal elements can be chosen to zero, $\chi_{ii}=0$, without loss of generality. Thus, there are a six independent entries in the interaction matrix:
\be
\doubleunderline{\chi} =  \;  \; \begin{blockarray}{cccccc}
    S& D & N & W  \\
\begin{block}{(cccc)cc}
     0 & \chi_{DS} & \chi_{NS} & \chi_{WS} &   \;\;\;\; S&\\
     \chi_{DS} & 0 & \chi_{DN} & \chi_{DW} & \;\;\;\; D&\\
     \chi_{NS} & \chi_{DN} & 0 & \chi_{NW} &  \;\;\;\; N\\
     \chi_{WS} & \chi_{DW} & \chi_{NW} & 0 &  \;\;\;\; W\\
\end{block}
\\
\end{blockarray} \qd
\ee

For the chemical reactions depicted in Fig.~\ref{fig:schemes}, we have $L=3$ reactions out of which $L'=2$ are linearly independent, see Eqs.~\eqref{eq:reac_1}-\eqref{eq:reac_2}.
The stoichiometric matrix of the linear independent reactions $\alpha=1$ and $\alpha=2$ is given by
\be
\doubleunderline{\sigma} = \;\;
\begin{blockarray}{cccc}
\alpha=1 & \alpha=2  \\
\begin{block}{(cc)cc}
    0  & 0   &   \;\;\;\; S&\\
    1  & -1   &   \;\;\;\; D&\\
    -1 & 0    &   \;\;\;\; N\\
    0  & 1   &   \;\;\;\; W\\
\end{block}
\\
\end{blockarray}\qd
\ee
Conserved densities $\psi_j$ and reaction extents $\xi_\alpha$ are obtained from volume fractions via the transformation matrix
\be
\doubleunderline{U} = \nu^{-1} \begin{pmatrix}
        1& 1&1&1 \\
       0& 1 & 1 & 1\\
       0& 0 & -1 & 0\\
       0&1 & 1 &2
        \end{pmatrix} \qc
\ee
see \eq{Umatrix}.
In addition to the volume conservation, the relevant conserved density $\psi=(\phi_D+\phi_N+\phi_W)/\nu$ describes the conservation of non-solvent components. The reaction extent variables are $\xi_1=-\phi_N/\nu$ and $\xi_2=(\phi_D+\phi_N-2\phi_W)/\nu$.
The inverse transformation matrix is given by
\be
\doubleunderline{U}^{-1} = \nu \begin{pmatrix}
        1&-1&0&0\\
        0&2 & 1 & -1\\
        0&0 & -1 & 0\\
        0&-1 & 0 &1
        \end{pmatrix} \qc
\ee
and obeys the equation $U^{-1}_{i\alpha}= \nu_i \sigma_{i \alpha}$.

Reaction rates are specified by kinetic coefficients $k_\alpha$ which depend on composition.
In our model,
the reactions $\alpha=1,2$ occur only inside the droplet phase with a rate
\be
k_\alpha = \frac{K_\alpha}{2}\left(1+\tanh\left(\frac{\phi_D-\Phi}{\epsilon}\right)\right) \qc
\ee
where $K_\alpha$ is a rate constant.
Moreover, $\Phi$ denotes a parameter corresponding to the value of the volume fraction $\phi_D$ at the interface between droplet phase and solvent phase, and $\epsilon>0$ characterizes how much the reaction $\alpha=1,2$ is suppressed outside the droplet.
In the bulk driven case, the reaction $\alpha=3$ occurs outside the droplet, corresponding to  $\epsilon<0$. In addition, the system is driven by an external energy input $\Delta \mu^{\text{act}}$. Therefore, we use for $\alpha=3$, instead of \eq{reactionrate_plusminus}, the following forward reaction rate
\be
\label{eq:mu_act_on_r3}
r_3^+=k_3(\phi_D) \exp\left(\frac{\mu_W + \Delta \mu^{\text{act}}}{k_BT} \right) \qc
\ee
and for rate of the backward reaction, we consider
\be
r_3^-=k_3(\phi_D) \exp\left(\frac{\mu_N }{k_BT} \right) \qd
\ee

\subsection{Steady state droplets}

\begin{figure*}[tb]
        \centering
        \includegraphics[width=0.99\textwidth]{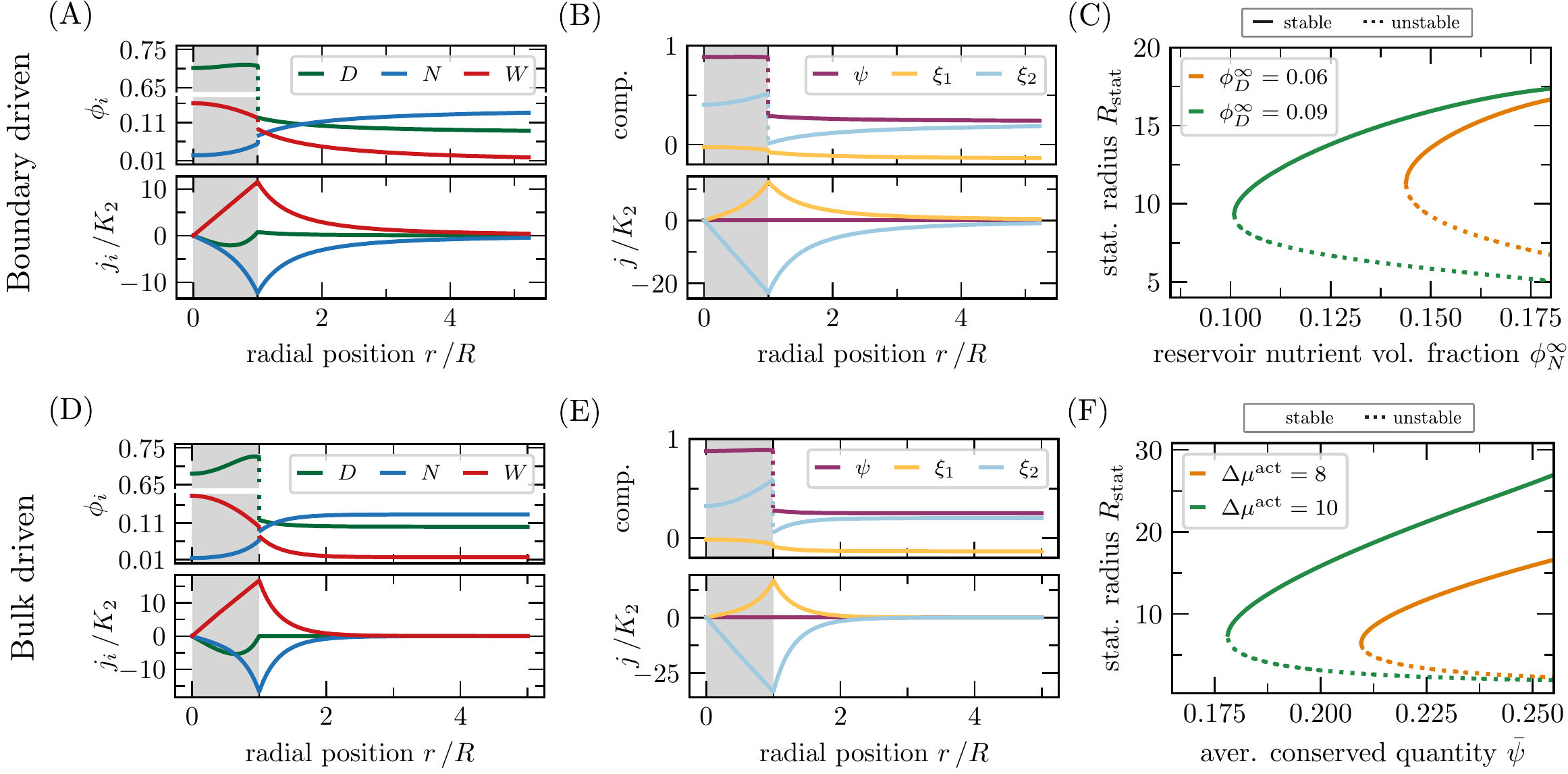} 
    \caption{
    \textbf{Stationary active droplets.}
    Profiles of various quantities
    for the boundary-driven case  (upper row) and the bulk-driven case (lower row).
    (A,D): Profiles of volume fractions and corresponding diffusive fluxes for droplet material, nutrient and waste.
    (B,E): Profiles of conserved density $\xi$ and non-conserved reaction extents $\xi_{1/2}$ and their corresponding diffusive fluxes. Such profiles are obtained from Eqs.~\eqref{eq:dyn_con} and \eqref{eq:dyn_ex}.
    (C,F): Stable and unstable stationary radii $R_\text{stat}$ of chemically active droplets.
    For the boundary-driven case, $R_\text{stat}$ is shown as a function of the nutrient volume fraction of the reservoir $\phi_N^\infty$ for two different reservoir values of the droplet material $\phi_D^\infty$.
    For the bulk-driven case,
    $R_\text{stat}$ is depicted as a function of the averaged conserved quantity $\bar{\psi}$ for two different values of external energy input $\Delta \mu^\text{act}$ (Eq.~\eqref{eq:mu_act_on_r3}) that break detailed balance of the rates~\eqref{eq:detailed_balance}.
    The parameters of these systems are given in Parameter values are given in Table~1 and Appendix~\ref{app:parameters}.
    }
    \label{fig:profiles_edm}
\end{figure*}

In order to characterize steady states of spherical droplets and their stability, we consider, for simplicity, the case of a sharp interface and an infinitely large system volume; see Appendix~\ref{app:effective_drop_model}.
In this limit of a sharp interface, we obtain linear reaction-diffusion equations~\eqref{eq:dyn_edm_lin} inside and outside the droplet.
These equations are derived by linearization around the concentrations that coexist at the interface which is at local equilibrium. We impose boundary conditions at the interface that ensure currents across the interface being continuous at steady state.
This continuity of currents selects a unique tie line which connects the coexisting concentrations in the phase diagram.

In the boundary driven case, we impose concentration boundary conditions at infinity with values $\phi_D^\infty$, $\phi_N^\infty$ and $\phi_W^\infty$ for the volume fractions of droplet material $D$, nutrient $N$ and waste $W$.
Fixed concentration boundary conditions imply that material is supplied via net currents at infinity.

Figure~\ref{fig:profiles_edm}(A) shows profiles of volume fractions of droplet material $D$, nutrient $N$ and waste $W$ in the boundary driven case for a spherical droplet of size $R$
as a function of the radial coordinate $r$.
The droplet material is produced by reaction $\alpha =1$ inside the droplet (grey region) where it is also the
majority component. The droplet material also occurs outside at low concentrations. Nutrient is provided at large distances and diffuse toward the droplet. Waste is produced inside the droplet by the reaction $\alpha=2$ and diffuses outwards.
The system is maintained out-of-equilibrium by the concentration boundary conditions at infinity, where $\mu_N \neq \mu_W$. Note that detailed balance is obeyed everywhere in the system.

Figure~\ref{fig:profiles_edm}(B) shows the profiles of the conserved density $\psi$ and the extent of the two reactions $\xi_1$, $\xi_2$ in the boundary driven case. The current of the conserved density vanishes, because the system is at steady state.
The divergence of the extent currents $\nabla \cdot j_\alpha^\xi = r_\alpha$ indicate chemical activity at rate $r_\alpha$. Note that outside the droplet, divergence-free extent currents occur without reactions.
The steady state droplet radius depends on the concentrations imposed at infinity, see Figure~\ref{fig:profiles_edm}(C), which shows stationary radii of stable (solid lines) and unstable (dashed lines) droplets for two different values of $\phi_D^\infty$ as a function of $\phi_N^\infty$.
Droplets nucleated beyond the critical radius (dashed lines) grow until the reach the stable stationary radius (solid lines), while droplets larger than the stable radius shrink.
For smaller volume fractions $\phi_D^\infty$ droplet material is lost by diffusion towards infinity, requiring larger volume fractions $\phi_N^\infty$ to maintain the droplet.

In the bulk driven case, we impose that no net current of the conserved quantity $\vect{j}^\psi$ exits at infinity. In addition, reaction $\alpha=3$ is imposed at local chemical equilibrium with $\Delta \mu^\text{act} =\mu_N-\mu_W$, where $\Delta \mu^\text{act}$ is the chemical free energy supplied by a fuel.
Note that, the reaction $\alpha=3$ occurs outside of the droplet to regenerate nutrient from waste. The profiles of volume fractions of a steady state droplet are shown in \fig{profiles_edm}(D), together with the corresponding currents.
The profiles of volume fraction are qualitatively similar to the boundary driven case shown in \fig{profiles_edm}(A). The main difference is that the currents decay more quickly and no net current remain at infinity.
The corresponding conserved density and reaction extent are shown in \fig{profiles_edm}(E).

The boundary driven case and the bulk driven  case give similar profiles for the conserved density $\psi$ and the reaction extents $\xi_1$ and $\xi_2$
(compare \fig{profiles_edm}(A,B)  to \fig{profiles_edm}(D,E)).
A key difference is that the  divergence of the extent currents, $\nabla \cdot j^\xi$, in the boundary driven case  vanishes outside the droplets, while for the bulk driven case, this divergence is non-zero outside, corresponding to the rate of the reaction that produces nutrient.
Note that the reaction in the bulk driven case cannot alter the amount of conserved material, therefore, the droplets growth is limited via the conserved material.

The stable and unstable radii of stationary droplets are shown in \fig{profiles_edm}{(F)} for the bulk driven case as function of the conserved quantity $\psi^\infty$ and for two different values of $\Delta \mu^\text{act}$. Increasing $\Delta \mu^\text{act}$ enables chemically active droplets for smaller values of the conserved quantity $\psi^\infty$ and leads to larger stable stationary droplets. This behaviour is similar to that the boundary driven case, see \fig{profiles_edm}{(C)}. Increasing $\Delta \mu^\text{act}$ in the bulk driven case has the same qualitative effect on the stationary radius, as an increase of nutrient supply via $\phi_N^\infty$ in the boundary driven case, compare  \fig{profiles_edm}{(C)} and \fig{profiles_edm}{(F)}.


\subsection{Droplet stability}

So far, we have considered stationary droplets of spherical shapes. Due to non-equilibrium conditions, chemically active droplets can also undergo a shape instability and take non-spherical shapes and thereby even divide. An example of a division event is shown in \fig{num}(A).
We can systematically study the linear stability of spherical shapes using the sharp interface limit, see Appendix~\ref{app:effective_drop_model}.


In \fig{stability}{(A)}, we show a stability diagram of chemically active droplets for the boundary driven case as function of nutrient and droplet material volume fraction at infinity, $\phi_N^\infty$ and $\phi_D^\infty$, for fixed $\phi_W^\infty$. 
Stationary spherical droplets are stable within the green region of the diagram. Within the the yellow region, a spherical harmonic deformation mode with $l=2$ is unstable. This corresponds to an elongation of the droplet shape.
Similarly in the red region, a spherical deformation mode with $l=3$ is also unstable.
In the white region, no stationary droplets exit.
The figure shows, that starting from a stationary stable droplet, division can typically be induced by increasing
the supply of either nutrient or droplet material.
The binodal line of phase coexistence is shown as a solid blue line.
In the absence of chemical reactions, droplets within the binodal region grow, while outside they shrink.
Chemical reactions also permit the existence of chemically active droplets outside of the binodal region,
where droplet material is constantly lost towards the reservoir.
Under such non-equilibrium conditions, the volume fractions at the interface (governed by the binodal) differ from the values at large distances imposed by boundary conditions.
The purple solid line indicates chemical equilibrium of reaction $\alpha=1$, where the chemical rate $r_\alpha$ changes sign. For nutrient volume fractions above this line, droplet material is produced by the nutrient, while below this line, nutrient is produced.

Figure~\ref{fig:stability}{(B)} shows the corresponding stability diagram for the bulk driven case as a function of the conserved density at infinity $\psi^\infty$ and the active chemical free energy $\Delta \mu^\text{act}$.
The same regions of stability are indicated: stable spherical droplets green, unstable mode with $l=2$ yellow and unstable mode with $l=3$ red.
Spherical stable droplets will typically divide when the conserved quantity is supplied, i.e.,  the conserved density $\psi^\infty$ is increased.
Moreover, for increasing values of $\Delta \mu^\text{act}$, the stability of active droplets becomes independent of $\Delta \mu^\text{act}$. This is because almost all waste is turned over to nutrient by
the chemical reaction $\alpha=3$.

\begin{figure}[tb]
        \centering
        \includegraphics[width=0.36\textwidth]{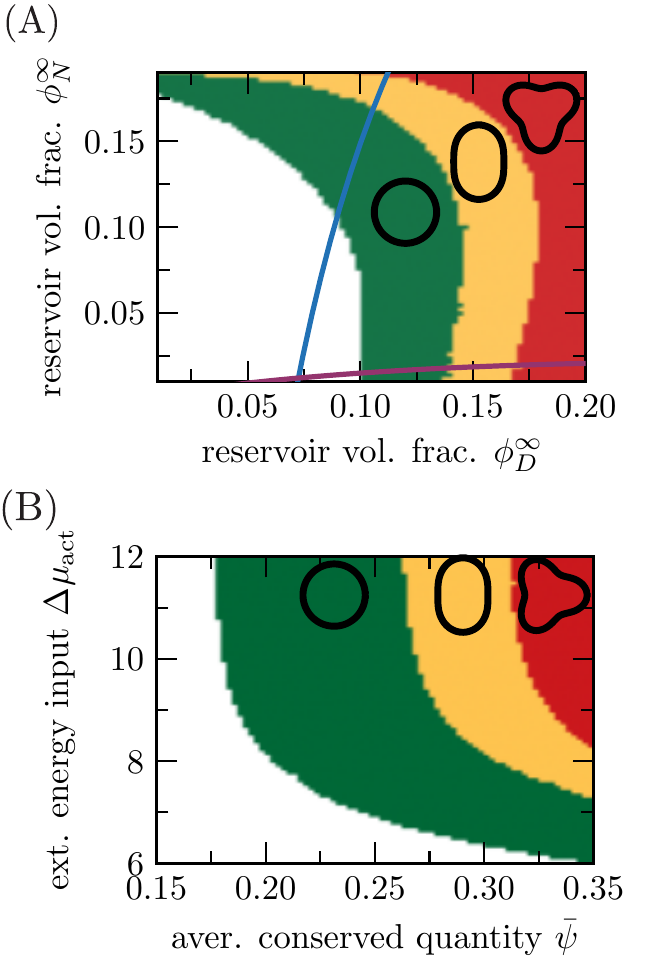} 
    \caption{
    \textbf{Shape stability diagrams for chemically active droplets} for the boundary driven case in (A) and the bulk driven case (B). In the green area stationary droplets of spherical shape can exist. All shape perturbations decay. In the yellow area, the slowest $l=2$ perturbation becomes unstable, thus droplets  elongate. In the red area, additionally the slowest $l=3$ mode become unstable.
    For the boundary driven case, we show the stability for a specific value of   waste volume fraction of the reservoir,  $\phi_W^\infty=0.001$. Therefore, we can draw the chemical equilibrium line of reaction $\alpha=1$ (purple) and the binodal line of these systems in the absence of chemical reactions (blue).
    Parameter values are given in Table~1 and Appendix~\ref{app:parameters}.
    }
    \label{fig:stability}
\end{figure}

\section{Energy, mass and entropy balance of protocells}\label{sect:energymassentropy_balance}
\begin{figure*}[tb]
        \centering
        \includegraphics[width=0.99\textwidth]{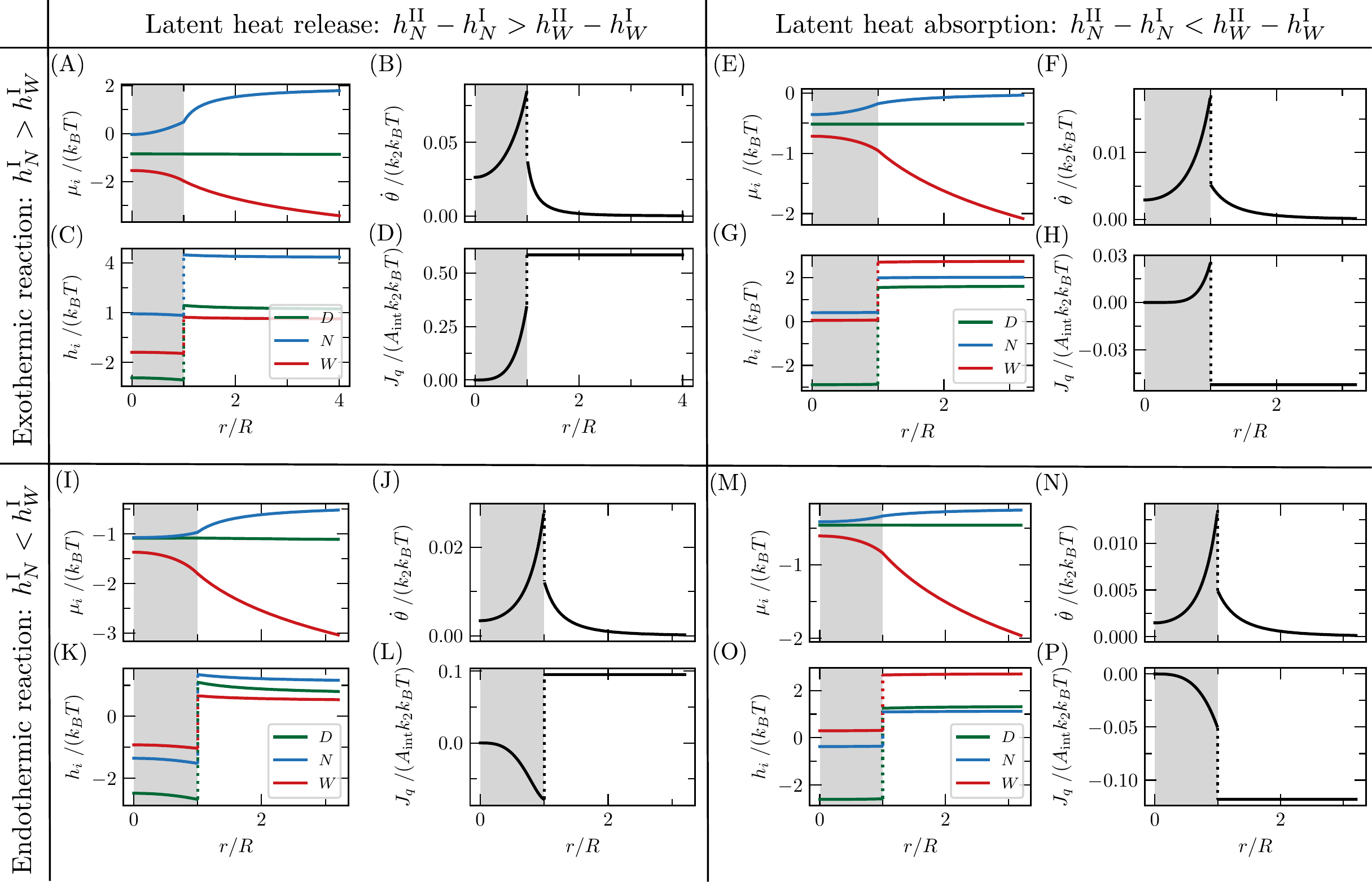} 
    \caption{
    \textbf{Energetics of active droplets.}
    We show the profiles of chemical potentials for four different boundary driven stationary active droplets in the panels (A,E,I,M). Due to the constant driving, these stationary droplets are non-equilibrium steady states characterized by their entropy production, shown in (B,F,J,N). Despite similar profiles of chemical potentials, these four scenarios differ in their profiles of molecular enthalpies $h_i$, shown in (C,G,K,O). Thus the total heat flux ${J}_q=\vect{e}_r \cdot \vect{J}_q$ varies respectively, see (D,H,L,P).
    For simplicity, we have chosen $\phi_D^\infty$ such that the current of droplet material  vanishes outside of the droplet in the steady state. Thus there is no flux of the droplet material over the interface at stationary state, which would contribute to latent heat production or absorption.
    Parameter values are given in Table~1 and Appendix~\ref{app:parameters}.}

    \label{fig:heat}
\end{figure*}

We now discuss energy and mass balance of stationary active droplets in the boundary driven case, where conserved quantities are only supplied at large distance.
The fluxes of the conserved quantity $\psi$ carry material associated with the growth and shrinkage of droplets. In a stationary case, these fluxes are zero.
Nevertheless, the chemical reactions within the droplet lead to gradients of the chemical potentials, $\nabla \mu_i \neq 0$.
At the same time, chemical reactions are also maintained out-of-equilibrium, corresponding to a non-vanishing Gibbs reaction free energy $ \Delta \mu_\alpha\neq0$.
Both, chemical and diffusive fluxes give rise to a production of entropy according to \eq{entropy}.

To discuss energy balance, we also consider heat.
Heat is released by chemical reactions if the reaction enthalpy $\Delta h_\alpha>0$ (exothermic), and is absorbed by chemical reactions if $\Delta h_\alpha<0$ (endothermic).
In addition heat is absorbed or released at the droplet interface (latent heat) for
$\sum_{i} \vect{j}_i \nabla h_i > 0$ and $\sum_{i} \vect{j}_i \nabla h_i < 0$, respectively. Here, the sum is over all solute species.

Figure~\ref{fig:heat}(A-P) present four scenarios which differ in the heat release by reactions and the heat release at the interface.
In all of the four scenarios, the imposed chemical potentials at the large distance start high for the nutrient, is lower for droplet material and lowest for waste, $\mu_N^\infty>\mu_D^\infty>\mu_W^\infty$, see chemical potential profiles in \fig{heat}(A,E,I,M).
This biases reactions $\alpha=1$ and $\alpha=2$ in the forward direction and diffusive transport and chemical reactions run in the same directions in all four scenarios.
Correspondingly, the radial profiles of local entropy production $\dot{\Theta}$ is very similar in all four scenarios, see \fig{heat}(B,F,J,N).
Note that the entropy production rate $\dot{\Theta}$ jumps at the interface because the reaction rates  $r_\alpha$ change discontinuously at the interface.
The maxima of entropy production rate occur close to the interface since newly supplied material first reaches the droplet surface where it starts to react.

The first scenario of exothermic reaction and latent heat release at the interface is shown in \fig{heat}(A-D).
In addition to the profiles of chemical potentials and entropy production, it is characterized by profiles of molecular enthalpies $h_i$ and total heat flux $\vect{J}_q=4 \pi \vect{j}_q$, see \fig{heat}(C-D).
The nutrient enthalpy inside the droplet $h_N^\I$ (blue) is larger than the waste enthalpy $h_W^\I$ (red), indicative of a net exothermic reaction from $N$ to $W$. In addition, \fig{heat}(C) shows that at the interface $h_N^\II-h_N^\I>h_W^\II-h_W^\I$
and, corresponding to a net latent heat release.
Here, the index I refers to droplet phase and II too the phase outside.
The release of heat at distance $r$ corresponds to the slope of total heat flux $d \vect{J}_q / dr$.
Heat release by chemical reactions corresponds to an  increase of $\vect{J}_q$ inside the droplet, the contribution of latent heat is captured by a discontinuity of $\vect{J}_q$ at the interface.

There are three further additional scenarios are shown in \fig{heat}:
Exothermal reactions with latent heat absorption at the interface \fig{heat}(E-H), endothermal reactions with heat release \fig{heat}(I-L) and endothermal reactions with heat absorption \fig{heat}(M-P).
In the case of endothermal reactions, the enthalpy of the waste is higher than that of the nutrient, see \fig{heat}(K,O)
In the case of latent heat absorption, $\vect{J}_q$ drops at the interface, such that the heat generated inside the droplet by exothermic reactions is not fully transported to the outside \fig{heat}(H).
Finally, in the case of endothermic reactions and heat absorption at the interface, the system takes up heat, which enters at large distances and is absorbed by reactions inside the droplet, \fig{heat}(P). The droplet, therefore, acts as a cooling device, which in the case of finite heat conductivity would lead to lower temperatures inside the droplet compared to the outside.

\section{Discussion}\label{sect:discussion}

By combining the physics of phase separation and the thermodynamics of chemical reactions governing the law of mass action, we provide a theoretical framework for
chemically active droplets.
Such droplets have been proposed as models for protocells which are prebiotic, cell-like objects that could have emerged at the origin of life~\cite{haldane:1929,oparin:1953,oparin:1959, hyman2012retrospect}.
Chemically active droplets rely on phase separation leading to coexisting phases which organise chemical reactions that are maintained away from equilibrium.
Using our framework, we discuss two cases how such droplets can be maintained away from equilibrium, see \fig{schemes}:
(A) Droplets coupled to external reservoirs at the system boundary supplying nutrient and removing waste. In this case, detailed balance holds in the system and non-equilibrium conditions only enter via the reservoirs at the boundaries.
(B) Droplets driven by a chemical fuel in the bulk. In this case, detailed balance of the chemical reaction between nutrient and waste is broken in the bulk.
Both cases (A) and (B) can exhibit non-equilibrium steady states with non-vanishing currents of energy and matter.

In our work, we discuss the balance of matter by introducing conserved densities and the non-conserved reaction extents.
In addition, we consider the balance of energy.
We show that the growth or shrinkage of a chemically active droplet is governed by fluxes of the conserved quantities at the droplet surface.
We further show that the maintenance of non-equilibrium steady states is enabled by fluxes of the non-conserved reaction extents.
From the point of view that active droplets represent simple models for protocells, the chemical reactions inside the droplets represent a simple metabolism.

In biology, metabolic processes are classified as anabolic or catabolic.
Anabolic processes are considered to build complex components from smaller units by consuming energy, while catabolism typically describes the break-up of complex molecules to smaller units by which energy is released.
The reactions in our models of protocells can capture such anabolic and catabolic processes.
As an example, we can consider nutrient and waste to be simple molecules of high and low internal energy, respectively. Droplet material $D$ would then represent more complex components. In such a setting, the reaction $N \rightleftharpoons D$ ($\alpha=1$) corresponds to anabolic   processes, while the reaction $D \rightleftharpoons W$ ($\alpha=2$) corresponds to catabolic processes.

Our work highlights the different roles of conserved densities $\psi_j$ and of non-conserved reaction extents $\xi_\alpha$ for protocell dynamics. This distinction between conserved and non-conserved densities is relevant not only for protocells, but also for biological cells.
We can distinguish processes associated with mass and volume growth from processes that maintain non-equilibrium conditions even in the absence of growth.
Growth processes in cells typically require the production of complex molecules by anabolic processes.
Maintaining a cell away from equilibrium requires the supply of chemical fuels, such as ATP, via catabolic processes.
Therefore, anabolism and catabolism can be related to growth and maintenance, respectively.
In our formalism describing protocells, growth is related to the accumulation of conserved densities $\psi_j$, while maintenance is associated with the dynamics of non-conserved reaction extents $\xi_\alpha$.
Our work shows that maintaining reaction extents away from equilibrium is a hallmark feature of living systems.

A signature of non-equilibrium state of living systems is the release of heat \cite{rodenfels:2019}. Our framework captures energy balance and heat exchange.
Nutrients supply free energy via a high chemical potential as compared to waste $\mu_N>\mu_W$, which drives the system out-of-equilibrium.
Typically, nutrient $N$ is also a molecular energy carrier and waste $W$ a low energy molecule which is reflected in molecular enthalpies ($h_N>h_W$).
In this case, active droplets are exothermic and release heat, see \fig{heat}(A-D).
However, protocells could also be endothermic if the molecular enthalpy of waste exceeds that of nutrient
($h_N<h_W$). In this case, the droplet would absorb heat from the environment, see \fig{heat}(M-P).
The latent heat of phase separation at the interface also enters this energy balance. Latent heat can be either released or absorbed, which leads to overall four different scenarios shown in \fig{heat}.
One of these scenarios shows that even if the droplet absorbs heat and appears to be endothermic, the reactions inside could still be exothermic, see \fig{heat}(E-H).
This shows that a measurment of overall heat absorption by an organism does not neccessarily imply endothermic biochemistry, see \cite{heijnen:1992,liu:2001}

Chemically active droplets can also divide.
Division of chemically active droplets was suggested in theoretical works for liquid droplets, either by modulating surface tension using chemical gradients \cite{rashevsky:1934}, or via a dynamical shape instability studied in a minimal model \cite{zwicker:2017}.
Here, we have discussed shape instabilities leading to droplet division in multi-component mixtures motivated by protocells.
In a binary system, chemical reactions have to convert droplet material $D$ and solvent $S$, which phase-separate, into each other.
We have shown that droplet division can occur in multi-component mixtures,
when the solvent $S$ that phase-separates from the droplet material $D$ is not taking part in chemical reactions.

Our work is relevant for experimental studies of active droplets. Such experimental systems typically involve multiple components that undergo chemical reactions~\cite{spath2021molecular,abbas2021peptide, schwarz2021parasitic, nakashima2021active}.
As our work shows, simply providing energy via a fuel is not sufficient to drive growth, because growth requires the accumulation of a conserved quantity.
Similarly, cycles of droplet growth and division also requires the supply of a conserved quantity.
These different roles of conserved densities for growth and non-conserved reaction extents for maintenance highlight the importance of energy and matter supply for chemically active droplets.

\appendix

\section{Free Energy}
\label{app:free_energy}

The chemical potentials given in \eq{chem_pot} are derived from the free energy
\be
F=\int d^3x \left[ f(\phi) +\sum_{i=1}^M \frac{\kappa_i}{2 \nu_i} (\nabla \phi_i)^2 \right]\qc
\label{eq:Total_free_energ}
\ee
with the Flory-Huggins type of free energy density
\be
 f=\sum_{i=0}^M\frac{k_B T}{\nu_i} \phi_i \log(\phi_i)+ \sum_{\langle i,j\rangle} \chi_{ij} \phi_i \phi_j + \sum_{i=0}^M\omega_i^0 \phi_i  \qc
\ee
where $\phi_0=1-\sum_{i=1}^M\phi_i$, and the sum over the pairs $\langle i,j \rangle$ include all possible combinations of different $i$ and $j$, including the solvent.
For simplicity, we have neglected contribution related to gradient in solvent volume fraction.
Moreover,  $\chi_{ij}$ is a matrix describing  the molecular interactions with  $\chi_{ii}=0$ and $\omega_i^0$ are the internal free energies.
After replacing the solvent volume fraction and using the definition $\mu_i=\nu_i \delta F/\delta \phi_i$, we obtain
\bg
\mu_i = k_B T \log \left( \phi_i \phi_0^{-\nu_i/\nu_0} \right) + 1-\frac{\nu_i}{\nu_0} \nn - \kappa_i \nabla^2 \phi_i  \\
+ \nu_i \sum_{j=1}^M  (\chi_{ij}-\chi_{i0}-\chi_{j0}) \phi_j
+ \nu_i\chi_{i0} + \nu_i \omega_i^0- \nu_i \omega_0^0  \qd
\eg
We can now identify the composition independent reference chemical potential
\be
\omega_i = 1- \frac{\nu_i}{\nu_0} + \nu_i\chi_{i0} + \nu_i \omega_i^0- \nu_i \omega_0^0 \qc
\ee
and the exchange activity coefficient as stated in \eq{act_coef}.

\section{Effective description of a single active droplet}\label{app:effective_drop_model}

The non-linear, fourth-order partial differential equations~\eqref{eq:dynamic} govern the spatial dynamics of concentration fields.
To analyze the shape stability of active droplets, we derive an effective description for the dynamics of a single active droplet for the limit  of a sharp  interface.
To this end, we linearize dynamic equations~\eqref{eq:dynamic} leading for each component to two linear, second-order partial differential equations coupled via a moving boundary condition at the droplet interface, which we refer to as interface conditions in the following.
For each time point,
such non-linear boundary conditions are solved by a set of volume fractions right inside and outside of the interface. These volume fractions serve as  linearization points for the dynamic equations.

For the boundary driven case, we can solve for the interface dynamics considering a quasi-stationary limit. In particular, we use the stationary solutions to the lineararized equations and can solve for the slow interface dynamics. This slow dynamics is driven divergence-free currents of the conserved quantities. In the bulk driven case, we solve directly for the stationary positions of the interface. Due to the absence of divergence-free currents, using stationary solutions for the differential equations for the volume fractions automatically enforce a resting interface.

\subsection{Spherical symmetric stationary droplet}

Linearizing \eq{dynamic} around the volume fractions at the interface $\Phi_i^\III$, where I and II indicate the droplet phase and phase outside of the droplet, respectively:
\bg
\partial_t \phi_i = \sum_j \left[ D_{ij}^\III\nabla^2 \phi_j + k_{ij}^\III \left(\phi_j-\Phi_j^\III\right) \right]+c_i^\III \label{eq:dyn_edm_lin},
\eg
where $D_{ij}^\III=D_{ij}(\Phi^\III)$ are the diffusion constants introduced in \eq{diff_const}.
Furthermore, linearizing the reaction rates $r_i$ given in \eq{reaction_rates} give the reaction coefficients
\be
k_{ij}^\III = \frac{\partial r_i(\Phi_i^\III)}{\partial \phi_j}
\ee
and constant source or sink terms $c_i=r_i(\Phi_i^\III)$.
This constant only vanishes if the interface values correspond to a chemical steady state with $r_i=0$.

We calculated stationary solutions to Eq.~\eqref{eq:dyn_edm_lin}  for each phases in the case of a single droplet of radius $R$ as a function of the volume fractions at the interface. The solutions for $\phi_i^\II$ have to obey the boundary conditions at $r\to\infty$. For the boundary driven case, concentrations of nutrient, waste and droplet material are imposed, while for the bulk-driven case, no flux boundary conditions are used for all components.
At the droplet center $r=0$, the flux vanishes for each components and both cases.
There are  $2M$ unknown volume fractions at the interface.
In addition, the interface velocity
\be v= \frac{dR}{dt}  \qc \ee
for the boundary-driven case, or the interface position $R$ for bulk-driven case has to be determined via interface conditions, respectively.
In total, this leads to  $(2M+1)$ unknowns.

Locally, at the droplet interface, we assume phase equilibrium.
From this, we obtain $M$
conditions from the balance of chemical potentials
\be
\mu_i^\I = \mu_i^\II \qc
\ee
and one condition from the balance of the osmotic pressures right inside and outside of the droplet~\cite{safran:2003, weber2019physics}:
\be
f^\I - f^\II = \sum_{i=1}^M \frac{\mu_i^\III}{\nu_i} \left(\Phi_i^\I - \Phi_i^\II \right) - \frac{2 \gamma(\Phi^\III)}{R} \qc
\ee
where $\gamma$ denotes the surface tension $\gamma$.
In multi-component mixtures,  the surface tension $\gamma$ varies for different phase equilibria along the binodal line.
The method developed by de Gennes~\cite{degennes:1985} can be generalized to multi-component mixtures for the case when $(\nabla \phi_D)^2$ is the dominant gradient contribution to the free energy~\eqref{eq:Total_free_energ}.

In systems without chemical reactions global, conservation laws of all $M$ components dictate the selection of phase equilibra.
In open systems, like in the considered boundary-driven case,  the local conservation laws at the moving interface select specific phase equilibria. The remaining $M$ conditions at the interface are given by
\be
v \, \vect{e}_r = \frac{\vect{j}_i^\I - \vect{j}_i^\II}{\Phi_i^\I - \Phi_i^\II} \cdot \vect{e}_r\qd \label{eq:drdt_open}
\ee
This equations holds for each component for a moving interface. Thus, in total, we obtain the $(2M+1)$ conditions at the interface.

In closed systems, like in our bulk-driven case with no-flux boundary conditions, local conservation laws at the interface of the reaction extents and global conservation laws of the conserved densities dictate the specific phase equilibria.
Furthermore, in general, the interface cannot move when only stationary solutions of \eq{dyn_edm_lin} are considered.
In the absence of divergence-free currents, all currents of the conserved densities vanish in a stationary state. Thus, when the conserved density has different values in the phases, the droplet cannot grow.

Let us demonstrate the consequence of \eq{drdt_open} in a ternary example with components $A$, $B$, the solvent $S$, and the chemical reaction $A \rightleftharpoons B$. In the stationary state for a closed systems $\vect{j}_A^\III =- \vect{j}_B^\III$. Using this relationship, and \eq{drdt_open}, we obtain
\be
\frac{\vect{j}_A^\I - \vect{j}_A^\II}{\Phi_A^\I - \Phi_A^\II} = -\frac{\vect{j}_A^\I - \vect{j}_A^\II}{\Phi_B^\I - \Phi_B^\II} \qc
\ee
which can either be true when $\vect{j}_A^\I=\vect{j}_A^\II$ for a resting interface, or
$\Phi_A^\I + \Phi_B^\I=\Phi_A^\II + \Phi_B^\I$
corresponding to a constant conserved density between the phases.
Therefore, we can only solve for stationary interface position in general. Here, the remaining $M$ equations consist out of $C$ global conservation laws equations of the conserved variables
\be
R^3 \psi_j^\I + (R^3_\text{sys}-R^3)\psi_j^\II = R^3_\text{sys} \bar{\psi_j} \qc
\ee
for finite systems with total system size $R_\text{sys}$.
In the limit of infinite systems, this simplifies to $\psi_j^\II=\bar{\psi}_j $.
From the local conservation laws of the $M-C$ reactions extents at the interface, we obtain
\be
\vect{j}_\alpha^{\xi,\I}=\vect{j}_\alpha^{\xi,\II}\qd
\ee
Therefore, we obtain again $2M+1$ equations, which determine the interface values and the stationary droplet radius $R$.

\subsection{Stability of spherical droplets}

\begin{table*}[tb]

    \begin{tabular}{l|c|c|c|c|c|c|c|c|c|c|c|c|c|c|c|c} 
 & $\chi_{NS}$&  $\chi_{WS}$& $\chi_{DN}$ &  $\chi_{DW}$    & $\omega_D^0$ & $\omega_N^0$ & $\omega_W^0$ & $K_1 e^{-\omega_N^0}$    &       $K_2 e^{-\omega_D^0}$ &       $K_3e^{-\Delta \mu^{\text{act}}}$ & $\phi_D^\infty$ & $\phi_N^\infty$ &$\phi_W^\infty$ & $\Delta \mu^{\text{act}}$ &$\bar{\psi}$  \\ \hline
Fig. 2 & 1&0&0&1&0& 10 & 20  & 0.025 &0.001  & - & 0.08 & - & 0.0001 & -& -\\ \hline
Fig. 3(A,B) &-0.5 &0.5 &0 &0  &3  &8 &0&0.3 & 0.0088 & -&0.08 & 0.15& 0.001& -&-\\ \hline
Fig. 3(C) &-0.5 &0.5 &0 &0  &3  &8 &0&0.3 & 0.0088 & -& - & -& 0.001& -&-\\ \hline
Fig. 3(D,E) &-0.5 &0.5 &0 &0  &3  &8&0 &0.3 & 0.0088 & 0.008&- & -& - & 9 & 0.25\\ \hline
Fig. 3(F) &-0.5 &0.5 &0 &0  &3  &8 &0&0.3 & 0.0088 & 0.008&- &  -& - & - & -\\ \hline
Fig. 4(A) &-0.5 &0.5 &0 &  0&3  &8 &0&0.3 &0.0088 &- &- &- &0.001 &- &- \\ \hline
Fig. 4(B) &-0.5 &0.5 &0 &  0&3  &8 &0&0.3 &0.0088 &0.008 &- &- &- &- &- \\ \hline
Fig. 5(A-D) &1&0 &-1 &0 & -0.8& 4&1 & 0.3&0.0088&- & 0.106& 0.079&0.0001 & -&-\\ \hline
Fig. 5(E-H) &-1&0.5 &0 &0  & -0.6& 3&2.5& 0.3&0.0088&- & 0.100& 0.107&0.0001 & -&-\\ \hline
Fig. 5(I-L) &1&0 &-1 &0 & -0.8& 1&1 & 0.3&0.0088&- & 0.119& 0.164&0.0001 & -&-\\ \hline
Fig. 5(M-P) &-1&0.5 &0 &0 & -0.6& 2&2.5 & 0.3&0.0088&- & 0.115& 0.182&0.0001 & -&-
\label{tab:para}
    \end{tabular}
    \caption{Parameter values used in calculations shown in Figs. 2-5.}
\end{table*}

In this section, we investigate the
stability of spherical active droplets upon shape perturbations and study the relaxation dynamics of such perturbations:
\bg
\phi^\III_i(r,\varphi,\vartheta,t) = \hat{\phi}^\III_i(r)+\delta \phi^\III_i(r,\varphi,\vartheta,t) \qc \nn \\
R(\varphi,\vartheta,t) =\hat{R}+\delta R (\varphi,\vartheta,t),
\eg
where $\hat{\phi}_i(r)$ are the volume fraction profiles of the stable solutions and $\hat{R}$ is the stable droplet radius.
We use the separation ansatz for the perturbations of the volume fraction values $\delta \phi^\III_i$, and the interface position $\delta R$ of the form
\bg
\delta \phi^\III_i(r,\varphi,\vartheta,t) = \sum_{n,l,m} \epsilon_{nlm} \varrho_{inl}^\III(r) Y_{lm}(\vartheta,\varphi)\text{e}^{\tau_{nl} t} \qc \label{eq:ansat_tern} \nn \\
\delta R(\varphi,\vartheta,t) = \sum_{n,l,m} \epsilon_{nlm} Y_{lm}(\vartheta,\varphi)\text{e}^{\tau_{nl} t} \qd
\eg
Here, the index $n=0,...,\infty$, is related to the radial coordinate $r$, the index $l=0,...,\infty$, is related to the polar angle $\varphi$, and the index $m=-l,...,l$ is related to the azimuthal angle $\vartheta$.
The amplitude of each mode is given by $\epsilon_{nlm}$, and its relaxation rate with $\tau_{nl}$. Furthermore, $Y_{lm}$ are the spherical harmonics.
The radial problem reduces to $M$ coupled Bessel equations with the  solution given by
\be
\varrho_{inl}^\III(r) = \sum_{k=1}^M A_k^\III \rho_{ikml}^\III \text{b}^\III_{l}(\lambda_{knl}^\III r) \qc
\ee
where the component specific weight $\rho_{ikml}^\III$ and the inverse length-scale $\lambda_{knl}^\III$ are obtained from the $M$ independent solution of the coupled problem for fixed $n,l,m$.
The functions $\text{b}^\III_{l}(x)$ are either modified spherical Bessel functions of first or second kind, according to the boundary conditions at $r=0$ or $r \rightarrow \infty$.
Boundary conditions at the interface fix the $2M$ unknown coefficients of $A_k^\III$ and the relaxation rate $\tau_{nl}$.
From the condition of phase equilibrium in linear order, we obtain
\bg
\sum_{j=1}^M \frac{\partial \mu_i^\I}{\partial \phi_j}
 \left( \partial_r \hat{\phi}_j^\I(\hat{R}) +  \varrho_{jnl}^\I(\hat{R})\right)= \nn \\
 \sum_{j=1}^M \frac{\partial \mu_i^\II}{\partial \phi_j}
 \left( \partial_r \hat{\phi}_j^\II(\hat{R}) +  \varrho_{jnl}^\II(\hat{R})\right) \qc \\
 \frac{\gamma(\Phi^\III) (l^2+l-2)}{\hat{R}^2}=\sum_{i=1}^M \frac{\phi_i^I-\phi_i^\II}{\nu_i} \nn\\
\left(
\sum_{j=1}^M \frac{\partial \mu_i^\III}{\partial \phi_j}
 \left( \partial_r \hat{\phi}_j^\III(\hat{R}) +  \varrho_{jnl}^\III(\hat{R})\right)
\right) \qc
 \eg
thus $M+1$ conditions.
The conservation law of each component at the interface, requires in linear order:
 \bg
 \tau_{nl} (\phi^\I_i-\phi^\II_i) = - \sum\limits_{j=1}^M \Big[
 D_{ij}^\I \left(\partial^2_r\hat{\phi}_j^\I(\hat{R}) + \partial_r \varrho_{jnl}^\I(\hat{R})\right) \nn \\
-
 D_{ij}^\II \left(\partial^2_r\hat{\phi}_j^\II(\hat{R}) + \partial_r \varrho_{jnl}^\II(\hat{R})\right)  \Big] \qd
 \eg
With this, we can obtain in total $2M+1$ conditions at the interface, fixing the coefficients $A_k^\III$ and the relaxation rate $\tau_{nl}$. When $\tau_{nl}>0$, any small perturbation of the corresponding mode exponentially growths in linear order.

\section{Parameter choices}\label{app:parameters}
Here, we give the numerical values of the parameters used for producing the figures.
For all figures, we have chosen $\nu=1$, $\omega_S^0=0$, $\omega_S^0=0$, $\kappa_D=1$, $\lambda=1$, $k_B T = 1$, $\chi_{DS}=3$ and $\chi_{NW}=0$. As explained in Appendix~B, $\kappa_N=\kappa_W=0$ in the cases of the sharp interface limit. For the numerical simulation of \fig{num}, $\kappa_N=\kappa_W=1$.
The remaining parameter vary for the different cases. Their values are given in Table~1.
For \fig{num}, we used a cubic lattice with $N=128$ grid points in each dimension.
The time points are given in terms of $K_2$.
\newpage
\bibliography{lib.bib}
\end{document}